\documentclass{article}

\usepackage{mathtools}
\usepackage[final]{neurips_2022}
\usepackage{amsmath}
\usepackage[linesnumbered,ruled]{algorithm2e}

\usepackage{wrapfig}

\usepackage[utf8]{inputenc} 
\usepackage[T1]{fontenc}    
\usepackage{hyperref}       
\usepackage{url}            
\usepackage{booktabs}       
\usepackage{amsfonts}       
\usepackage{nicefrac}       
\usepackage{microtype}      
\usepackage{xcolor}         
\usepackage{amsmath,amssymb,amsthm,amsfonts}
\usepackage{latexsym,bbm,graphicx,float,mathtools}
\usepackage{cleveref}
\usepackage{caption}
\usepackage{subcaption}
\usepackage{transparent}

\newcommand{\dpdegree}{D_{\textsc{dp}}}
\newcommand{\ppdegree}{D_{\textsc{pp}}}
\newcommand{\tpdegree}{D_{\textsc{tp}}}

\newcommand{\dpbytes}{c_{\textsc{dp}}}
\newcommand{\ppbytes}{c_{\textsc{pp}}}
\newcommand{\coG}{{\mathbf{\widehat{G}}}}

\usepackage[inline]{enumitem}
\title{Decentralized Training of Foundation Models in Heterogeneous Environments}

\author{
Binhang Yuan$^{1}$\thanks{Equal contribution.}, Yongjun He$^{1}$\footnotemark[1], Jared Quincy Davis$^2$, Tianyi Zhang$^2$, Tri Dao$^2$,\\ 
\textbf{Beidi Chen}$^{34}$, \textbf{Percy Liang}$^2$, \textbf{Christopher Re}$^2$, \textbf{Ce Zhang}$^1$\\
    $^1$ETH Z\"urich, Switzerland~~~$^2$Stanford University, USA~~~$^3$Carnegie Mellon University~~~$^4$Meta AI\\
\{binhang.yuan, yongjun.he, ce.zhang\}@inf.ethz.ch\\
\{jaredq, tz58, trid, beidic, pliang, chrismre\}@stanford.edu
}

\begin{document}

\maketitle

\begin{abstract}
Training foundation models, such as GPT-3 and PaLM, can be extremely expensive, often involving tens of thousands of GPUs running continuously for months. These models are typically trained in specialized clusters featuring fast, homogeneous interconnects and using carefully designed software systems that support both data parallelism 
and model/pipeline parallelism. Such dedicated clusters can be costly and difficult to obtain. \textit{Can we instead leverage the much greater amount of decentralized, heterogeneous, and lower-bandwidth interconnected compute?} Previous works examining the heterogeneous, decentralized setting focus on relatively small models that can be trained in a purely data parallel manner. State-of-the-art schemes for model parallel foundation model training, such as Megatron and Deepspeed, only consider the homogeneous data center setting. In this paper, we present the first study of training large foundation models with model parallelism in a decentralized regime over a heterogeneous network. Our key technical contribution is a scheduling algorithm that allocates different computational ``tasklets'' in the training of foundation models to a group of decentralized GPU devices connected by a slow heterogeneous network. We provide a formal cost model and further propose an efficient evolutionary algorithm to find the optimal allocation strategy. We conduct extensive experiments that represent different scenarios for learning over geo-distributed devices simulated using real-world network measurements.
In the most extreme case, across 8 different cities spanning 3 continents, our approach is 4.8$\times$ faster than prior state-of-the-art training systems.
\end{abstract}

\section{Introduction}


Recent years have witnessed 
the rapid development of 
deep learning models, particularly foundation models (FMs)~\cite{bommasani2021opportunities} such as GPT-3~\cite{brown2020language} and PaLM~\cite{chowdhery2022palm}.
Along with these rapid advancements,
however, comes computational 
challenges in training these models: 
the training
of these FMs can be 
very expensive --- a single GPT3-175B training run takes 
3.6K Petaflops-days~\cite{brown2020language}--- this amounts to 
\$4M on today's AWS on demand instances,
even assuming 50\% device utilization
(V100 GPUs peak at 125 TeraFLOPS)! 
Even the smaller 
scale language models, e.g., GPT3-1.3B (1.3 billion parameters), on which this paper evaluates, require 64 Tesla V100 GPUs to run for one week, costing \$32K on AWS. 
As a result, speeding up training and decreasing the cost 
of FMs have been active research areas.
Due to their vast 
number of model parameters,
state-of-the-art systems 
(e.g., Megatron\cite{shoeybi2019megatron}, Deepspeed\cite{rasley2020deepspeed}, Fairscale\cite{baines2021fairscale})
leverage multiple forms of 
parallelism~\cite{shoeybi2019megatron,xu2021gspmd, zheng2022alpa,li2021terapipe,
rajbhandari2020zero,ren2021zero}.
However, their design is only tailored to 
\textit{fast}, \textit{homogeneous} data center networks.

\begin{figure}[t!]
\centering
\includegraphics[width=0.99\textwidth]{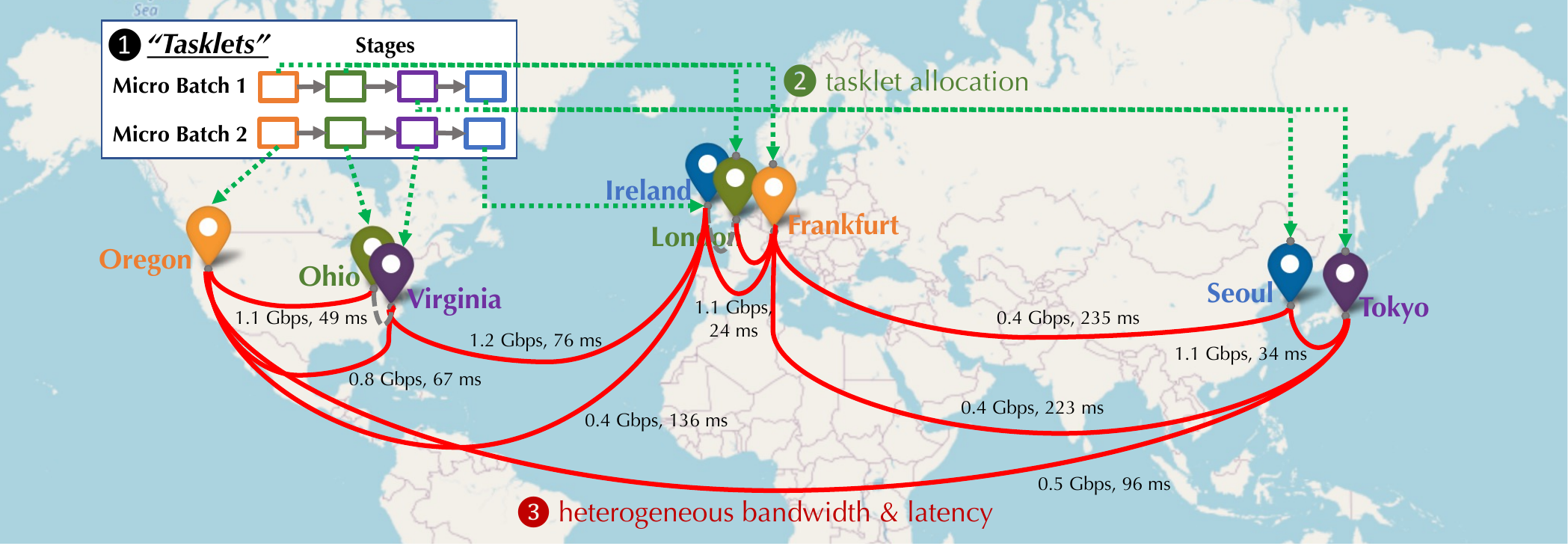}
\caption{Given \textcircled{1} a set of computation tasklets involved in training foundation models (corresponding to different micro-batches and layers), and \textcircled{2} a heterogeneous network between devices, the goal 
is to find the optimal \textcircled{3} allocation 
of  tasklets to devices.
}
\label{fig:geomap}
\end{figure}

On the other hand, \textit{decentralization} 
is a natural and promising 
direction. Jon Peddie Research reports that the PC and AIB GPU market shipped 101 million units in Q4 2021 alone~\cite{gpu_increase}. Furthermore, many of these GPUs are underutilized. Leveraging this fact, volunteer computing projects such as Folding@Home~\cite{shirts2000screen} have sourced upwards of 40K Nvidia and AMD GPUs continuously \cite{folding_stats}. Moreover, the incremental electricity and HVAC costs of running a V100 GPU 
for a volunteer are 50--100$\times$ lower than the spot prices for an equivalent device on AWS~\cite{economic_mining}.
If we could make use of these devices in a decentralized open-volunteering paradigm for foundation model training, this would be a revolutionary alternative to the expensive solutions offered by data centers.

This vision inspired 
many recent efforts
in decentralized learning, including 
both those that are theoretical and algorithmic~\cite{lian2017can, koloskova2019decentralized,koloskova2020unified}, as well as 
recent prototypes such as
Learning@Home~\cite{ryabinin2020towards} and DeDLOC~\cite{diskin2021distributed}.
However, efforts to-date in decentralized training either focus solely on \textit{data parallelism}~\cite{lian2017can, koloskova2019decentralized,koloskova2020unified,diskin2021distributed}, which alone is insufficient for FMs whose parameters exceed the capacity of a single device, or orient around alternative architectures, e.g., mixture of experts~\cite{ryabinin2020towards}. These alternative architectures
provide promising directions for 
decentralized learning, however, they are currently only trained and evaluated 
on smaller datasets and at a smaller computational scale (e.g., MNIST and WikiText-2 in~\cite{ryabinin2020towards})
than their state-of-the-art counterparts, e.g.,  GLaM~\cite{glam}.
In this paper, we focus on a standard 
GPT-style architecture, without considering
any changes that might 
alter the model architecture or the convergence behaviour
during training. 

To fulfill the potential of decentralization
for the training of FMs, 
we need to be able to (1) take advantage of 
computational devices 
connected via heterogeneous networks
with limited bandwidth and significant 
latency, and (2) support forms of parallelism
beyond pure data parallelism.
In this paper, we tackle one 
fundamental aspect of this goal ---
how can we assign different computational 
``tasklets'', corresponding to 
a micro-batch and a subset of layers,
to a collection of geo-distributed devices connected
via heterogeneous, slow networks?
This is not an easy task --- even for
fast and homogeneous data center networks,
such assignments are still an open
ongoing research challenge~\cite{narayanan2019pipedream, tarnawski2020efficient, tarnawski2021piper, fan2021dapple, park2020hetpipe}.
For the heterogeneous setting, it becomes even more challenging as the 
size of the search space increases 
dramatically.
In the 
homogeneous setting,
the homogeneity of the edges
in the communication graph 
reduces the search space into 
many equivalent classes representing
allocation strategies with the same communication costs, 
enabling 
efficient polynomial runtime
algorithms~\cite{tarnawski2020efficient,tarnawski2021piper,narayanan2019pipedream,fan2021dapple,park2020hetpipe}; however, in the heterogeneous setting, one has to consider 
potentially exponentially many
more distinct allocation 
strategies --- as we will see later, because of the heterogeneity of the communication matrix, even the sub-problem of finding the 
best pipeline parallelism strategy
equates to a hard open loop travelling salesman problem~\cite{papadimitriou1977euclidean}.

In this paper, we focus on this challenging 
scheduling problem of decentralized training
of FMs over slow, heterogeneous 
networks, and make the following contributions:

\begin{itemize}[leftmargin=2em,nosep,nolistsep]
\item We study the problem of allocating distributed training jobs over a group of decentralized GPU devices connected via a slow heterogeneous network. More specifically:
\begin{itemize}[leftmargin=2em,nosep,nolistsep]
    \item To capture the complex communication cost for training FMs,
    we propose a natural, but novel, formulation 
    involving decomposing the \textit{cost model} into 
    two levels: the first level is a \textit{balanced graph partitioning} problem corresponding to 
the communication cost of 
data parallelism, whereas the 
second level is a joint \textit{graph matching} and \textit{traveling salesman} problem corresponding to 
the communication cost of 
pipeline parallelism.
    \item We propose a novel \textit{ scheduling algorithm} to search for the optimal allocation strategy 
    given our cost model.
    Developing a direct solution to this optimization 
    problem is hard; thus,
    we propose an efficient 
    evolutionary algorithm based on 
    a collection of novel heuristics,
    going beyond the traditional 
    heuristics used in standard
    graph partitioning methods~\cite{kang2000hybrid}.
\end{itemize}
\item We carefully designed and 
    implemented a collection of \textit{system optimizations} 
    to hide communication within the computation to further reduce the impact of slow connections.\footnote{Our code is available at: \url{https://github.com/DS3Lab/DT-FM}.}
 
\item We conduct extensive experiments that represent different scenarios of collaborative decentralized learning, simulated by using network measurements from   
different geographical regions of AWS.
In the worldwide setting with 64 GPUs across 8 regions (Oregon, Virginia, Ohio, Tokyo, Seoul, London, Frankfurt, Ireland), 
we show that our system is 3.8-4.8$\times$ faster, in end-to-end runtime, than the state-of-the-art systems, for training GPT3-1.3B, \textit{without any difference in what is computed or convergence dynamics}.
In addition, we also provide careful 
ablation studies to show the individual effectiveness of the scheduler and system optimizations.

\item We shed light on the potential of decentralized learning --- our prototype in the 
global heterogeneous setting
is only \textit{1.7-3.5$\times$ slower than Megatron/Deepspeed in data centers even though its network can be 100$\times$ slower}.
We hope this paper can inspire future explorations of
decentralized learning 
for FMs, over geo-distributed
servers, desktops, laptops, or even mobile devices.
\end{itemize}

{\bf Limitations and Moving Forward.}
In this paper, we tackle one foundational 
aspect of decentralized learning but leave 
as future work
many problems that are important for a practical system. We assume that communication between devices is relatively stable for a reasonable amount of time and that all devices are always online without failure or eviction. Note that we also do not train a full system to full convergence, instead running partial training to confirm intermediate result equivalence across regimes.
Scheduling over a dynamic, heterogeneous environment and providing fault tolerance, potentially with checkpointing, while training to convergence are directions
for future exploration.

\begin{wrapfigure}{r}{0.375\textwidth}
  \begin{center}
  \vspace{-2.5em}
    \includegraphics[width=0.375\textwidth]{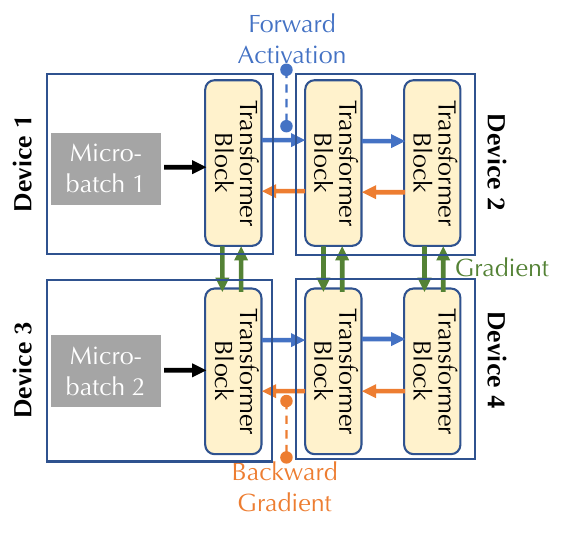}
  \end{center}
  \vspace{-2.5em}
\end{wrapfigure}

\vspace{-0.5em}
\section{Decentralized Training of Foundation Models: Problem Formulation}
\vspace{-0.5em}

We first introduce concepts, technical terms, and the procedure of decentralized training. Then we formally define the scheduling problem this paper tackles.

\textbf{Decentralized setting.} We assume a group of devices (GPUs) participating in collaborative training of a foundation model. Each pair of devices has a connection with potentially different delay and bandwidth. These devices can be geo-distributed, as illustrated in Figure~\ref{fig:geomap}, with vastly different pairwise communication bandwidth and latency. In decentralized training, all layers of a model are split into multiple \textit{stages}, where each device handles a consecutive sequence of layers, e.g., several transformer blocks~\cite{vaswani2017attention}. In addition, since the input for foundation model pre-training is huge, e.g., a few millions of tokens, it is also split into multiple \textit{micro-batches} that can be handled in parallel.

\textbf{Problem definition.} We define \textit{tasklets} as a collection of computational tasks in foundation model training --- Tasklet $t_{i, j}$ is the forward and backward computation for a 
stage $j$ with a micro-batch $i$ of training data in a training iteration. We aim to design an effective scheduler to assign each tasklet to a particular device so that the training throughput is maximized in decentralized training. 

\textbf{Parallelism.} The above setting involves two forms of parallelism, \textit{pipeline} and \textit{data}. In 
pipeline parallelism, the compute in multiple stages is parallelized --- each device handles activation or gradient computation for different micro-batches in parallel and the results can be communicated or passed to subsequent stages. Data parallelism means that devices compute the gradient for different micro-batches independently, but need to synchronize these gradients through communication. In a decentralized environment, the training procedure is \textit{communication-bounded}. The scheduling problem is to accelerate the communication procedure by allocating tasklets that require high communication volumes between them to devices with faster connections.

\begin{figure}[t!]
    \centering
    \includegraphics[width=0.975\textwidth]{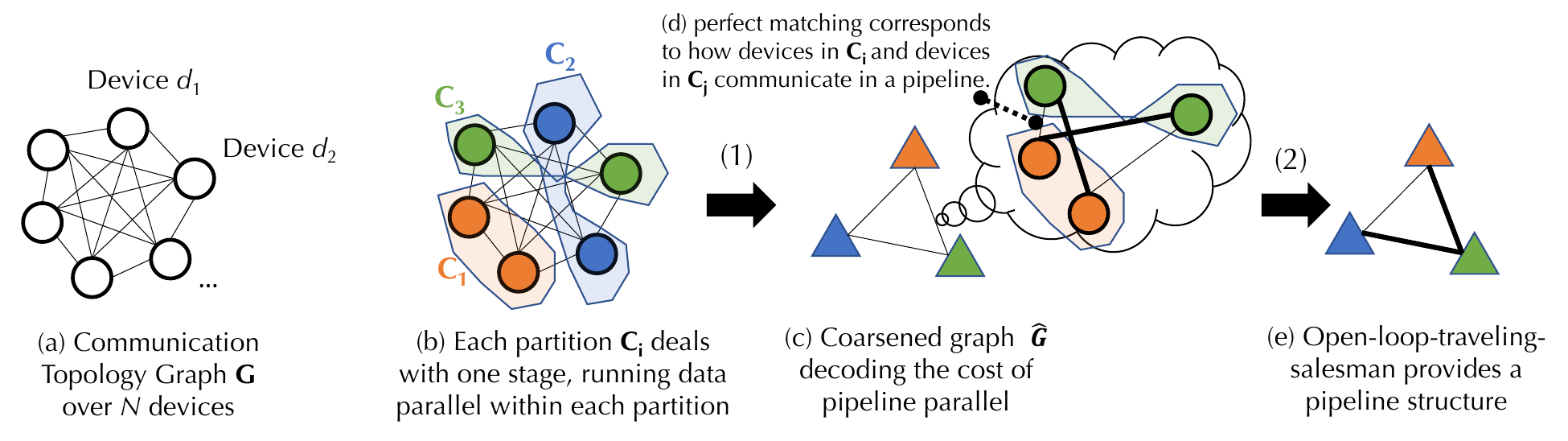}
    \vspace{-0.5em}
    \caption{(a) Communication graph $\mathbf{G}$; and (b, c, d, e) an illustration of the cost model given $\mathbf{G}$.}
    \label{fig:1.1}
    \vspace{-1.5em}
\end{figure}

\textbf{Formalization of the scheduling problem.} 
Formally, our scheduling problem is as follows.
\begin{itemize}[leftmargin=2em,nosep,nolistsep]
\item Let $\mathbf{D} = \{d_1 \ldots d_N\}$ be a set of $N$ devices;
$\mathbf{A} \in \mathbb{R}_+^{N\times N}$ and $\mathbf{B} \in \mathbb{R}_+^{N \times N}$ be the communication matrix between these devices describing the delay and bandwidth respectively, where the delay and bandwidth between device $d$ and $d'$ is $\alpha_{d,d'}$ and $\beta_{d,d'}$. 
\item Given the communication matrix 
$\mathbf{A}$ and $\mathbf{B}$, we construct a communication graph $\mathbf{G}$ (Figure~\ref{fig:1.1}(a)) --- each device corresponds to a node in $\mathbf{G}$ and each edge between
$d$ and $d'$ is labeled
with the average latency and bandwidth between $d$ and $d'$: $((\alpha_{d, d'} + \alpha_{d', d})/2, (\beta_{d, d'} + \beta_{d', d})/2)$.
Even though $\mathbf{A}$ and $\mathbf{B}$
are asymmetric (i.e., upload and download speeds might be different),
the communication graph $\mathbf{G}$ is symmetric 
because in our workloads all communications between 
two devices happen to involve the same amount of upload and download.
\item The number of stages that a micro-batch needs to go through is $\ppdegree$ (noted as pipeline parallel degree); the number of batch partition that needs to run model gradient synchronization is $\dpdegree$ (noted as data parallel degree); we have $\dpdegree \times \ppdegree = N$, i.e., the total number of devices.
\item $\ppbytes$ (resp. $\dpbytes$) represent the number of bytes of activations for a micro-batch (resp. parameters/gradients for a stage) communicated in pipeline parallelism (resp. data parallelism).

\item We denote a training tasklet as $t_{i,j}$, where $i \in \{1,..., \dpdegree\}$ and $j \in \{1,..., \ppdegree\}$, each of which corresponds to one specific micro-batch $i$ and
pipeline stage $j$.
\item An assignment strategy 
$\sigma \in \mathbf{D}^{D_{\text{DP}} \times D_{\text{PP}}}$
assigns,
for each tasklet $t_{i, j}$, a device
$\sigma_{i, j} \in \mathbf{D}$,
which means that device $\sigma_{i, j}$ runs the training tasklet $t_{i, j}$.
A valid assignment needs to be unique, i.e., $\forall (i, j) \ne (i', j')$: $\sigma_{i, j} \ne \sigma_{i', j'}$.
We use $\Sigma$ to denote the set of all 
valid assignments.
\item An \textit{optimal assignment strategy} is an assignment $\sigma$ that minimizes the
communication cost

\[
  \sigma^* = \arg\min_{\sigma \in \Sigma} \textsc{Comm-Cost}\left(\sigma\right)
\]
\end{itemize}

\textbf{Challenges and Goals.}
Our goal is to find 
the optimal assignment strategy, which involves
two challenges: (1) How to effectively model 
the communication cost 
$\textsc{Comm-Cost}(\sigma)$
for a given assignment $\sigma$
under a heterogeneous network environment?
and (2) How to effectively search 
for the optimal assignment strategy 
that minimizes such a cost?
We tackle these two questions in Section~\ref{sec:schedule}.

\vspace{-0.5em}
\section{Scheduling in Heterogeneous Environments}
\label{sec:schedule}
\vspace{-0.5em}

Scheduling in the heterogeneous setting is a challenging task, as the size of the search space increases dramatically compared to that of the homogeneous case.
In the homogeneous data-center 
case, the network delay can be usually ignored (e.g., $\mathbf{A}=\mathbf{0}$) and the bandwidth $\mathbf{B}$ are assumed to be formed by just a few constants --- e.g., the 
communication bandwidths between 
different machines on the same rack
are assumed to be same~\cite{tarnawski2020efficient,tarnawski2021piper,narayanan2019pipedream,narayanan2019pipedream,park2020hetpipe}. This significantly constrains the search space --- one can ignore the influence of communication given uniform connections~\cite{tarnawski2020efficient,tarnawski2021piper,narayanan2019pipedream}, or organize the device with a hierarchical structure~\cite{narayanan2019pipedream,park2020hetpipe}, making the scheduling problem solvable in polynomial time in terms of the number of machines.

In contrast, in the fully heterogeneous scenario the communication matrix 
$\mathbf{A}$ and $\mathbf{B}$ consists of 
distinct values, which can make the search space grows exponentially. In this section, we describe our scheduler that searches for an optimal strategy in the complex search space.

\subsection{Overview of the scheduler}

We carefully design a bi-level scheduling algorithm based on extended \textit{balanced graph partition} problem (see Figure \ref{fig:1.1}), and solve this problem by an evolutionary algorithm with a carefully designed local search strategy. 
Given an assignment strategy 
$\sigma = \{\sigma_{i, j}\}$
for all tasklets $\{t_{i, j}\}$,
we first model its
communication cost.
During the training of
FMs,
the communication costs come
from two different sources:
\underline{\textit{(1) Data parallel}:} All devices that are assigned 
with the tasklets dealing with the same stage $j$ (handling the same layers) of different micro-batches need to communicate within themselves to exchange gradients of these layers.
For layer $j$, we call these devices 
its \textit{data parallel group}:
$\mathbf{C}_{j} = \{\sigma_{i,j} ~|~ \forall i \in \left[ \dpdegree \right] \}$.
We can implement the communication using different primitives, e.g.,  \texttt{AllReduce}~\cite{sergeev2018horovod}, \texttt{ScatterGatter}~\cite{jiang2020unified},
or other decentralized average protocols~\cite{lian2017can}.
\underline{\textit{(2) Pipeline parallel}:} All devices that are assigned 
with the tasklets dealing with the same micro-batch $i$ of different
stages need to form a pipeline, communicating within themselves 
to exchange activations and backward gradients. For micro-batch $i$, these
devices are $\mathbf{P}_{i} = \{\sigma_{i,j} ~|~ \forall j \in [\ppdegree]\}$. Because these devices need to form 
a linear pipeline, any \textit{permutation} over 
$\mathbf{P}_{i}$ corresponds to 
one strategy of how these machines
can conduct pipeline parallelism within them.

\textbf{Scheduling Problem.}
The goal of our scheduler is to 
minimize both costs. 
One design decision that we made is
to decompose this complex optimization problem into two levels. 
At the first level, we
consider the best way of 
forming $\mathbf{C}_{j}$'s, incurring
data parallel communication 
costs within them. 
At the second level, 
we consider the cost of pipeline parallelism 
\textit{given} an layout from the first level:

\begin{equation}
\begin{aligned} 
\min_{\mathbf{C}_1 ... \mathbf{C}_{\ppdegree}} 
\textsc{Comm-Cost}\left(\mathbf{C}_1 ... \mathbf{C}_{\ppdegree}\right) :=&
\textsc{DataP-Cost}(\mathbf{C}_1 ... \mathbf{C}_{\ppdegree}) \\
&+
\textsc{PipelineP-Cost}(\mathbf{C}_1 ... \mathbf{C}_{\ppdegree})
\\
 s.t. \quad |\mathbf{C}_1| = .... = |\mathbf{C}_{\ppdegree}| = \dpdegree,
      &\forall j, j': \mathbf{C}_j \cap \mathbf{C}_{j'} = \emptyset, \mathbf{C}_1 \cup ... \cup \mathbf{C}_{\ppdegree} = \mathbf{D}
\end{aligned}
\label{eq:main}
\end{equation}

where computing 
$\textsc{PipelineP-Cost}(\mathbf{C}_1 ... \mathbf{C}_{\ppdegree})$
involves finding the optimal 
pipeline structure.

In Section~\ref{cost:dp} and Section~\ref{cost:pp}, we provide details on  $\textsc{Comm-Cost}\left(\mathbf{C}_1 ... \mathbf{C}_{\ppdegree}\right)$. 
Notice that this modified objective makes our problem different from the textbook graph partition problem; thus, we need a carefully designed evolutionary algorithm for finding such a solution introduced in Section~\ref{evol}. 

\subsection{Modelling data parallel communication cost} 
\label{cost:dp}

Given the communication graph $\mathbf{G}$
forming data parallel groups
$\mathbf{C}_1 ... \mathbf{C}_{\ppdegree}$
corresponds to 
a \textit{partition}
of $\mathbf{G}$---
In Figure~\ref{fig:1.1}(b), different
colors correspond to 
devices in the same 
$\mathbf{C}_j$.
The data parallel cost within
$\mathbf{C}_j$ only relies on 
all communication channels 
(edges in the communication graph)
connecting devices in $\mathbf{C}_j$. 
If we assume a colocated sharded parameter server~\cite{jiang2020unified} implementation for communicating within each $\mathbf{C}_j$,
and recall that $\dpbytes$ represents the
total amount of data (in bytes) that needs to be exchanged during gradient aggregation --- each device
in $\mathbf{C}_j$ needs to manage ${\dpbytes}/{\dpdegree}$ bytes of parameter shard. Once the gradient is ready, each device needs to send each of its local shards to the corresponding device; next, each device can aggregate the gradients it receives from all other devices in $\mathbf{C}_j$; and finally, each device will send the aggregated gradient shard to all other devices. Therefore, we can model the data parallel cost for $\mathbf{C}_j$ as follows:

\begin{equation}
\begin{aligned} 
\textsc{DataP-Cost}(\mathbf{C}_j)
=
\max_{d \in \mathbf{C}_j}
\sum_{
d' \in \mathbf{C}_j - \{d\}
}
2\cdot \left(\alpha_{d, d'} + \frac{c_{\text{dp}}}{\dpdegree {\beta}_{d, d'}}\right).
\end{aligned}
\end{equation}

\noindent Here, the total cost 
is bounded by the \textit{slowest} 
device ($\max_{d \in \mathbf{C}_j}$), which 
needs to exchange 
data with all other machines ($\sum_{d' \in \mathbf{C}_j - \{d\}}$).
Because the communication of these different data parallel
groups $\mathbf{C}_1 ... \mathbf{C}_{\ppdegree}$ can be conducted in parallel and we are
only bounded by the slowest data parallel group. This allows us to model the total communication cost for data parallelism as:

\[
\textsc{DataP-Cost}(\mathbf{C}_1 ... \mathbf{C}_{\ppdegree}) = \max_{j \in \left[\ppdegree\right]} \textsc{DataP-Cost}(\mathbf{C}_j)
\]

\subsection{Modeling pipeline parallel communication cost} 
\label{cost:pp}

Given $\mathbf{C}_1 ... \mathbf{C}_{\ppdegree}$,
to model the communication cost of 
pipeline parallelism, we need to consider two factors: (1) each \textit{permutation} $\pi$ of 
\{$\mathbf{C}_1 ... \mathbf{C}_{\ppdegree}$\}
corresponds to a specific pipeline strategy ---
devices in $\mathbf{C}_{\pi_{j}}$ and devices in $\mathbf{C}_{\pi_{j+1}}$ communicates to exchange 
activations (during forward pass) and gradients on activations (during backward pass); and
(2) devices in $\mathbf{C}_{\pi_{j}}$ and devices in 
$\mathbf{C}_{\pi_{j+1}}$ need to be ``matched'' --- only devices that are dealing with the same
micro-batch needs to communicate.
This makes modeling the cost of 
pipeline parallel communication more complex.

To model the cost of pipeline parallel communication, we first consider the best 
possible way that devices in 
$\mathbf{C}_j$ and $\mathbf{C}_{j'}$
can be matched. We do this
by creating a \textit{coarsened communication graph} (Figure~\ref{fig:1.1}(c)). 
A coarsened communication graph $\coG _{\mathbf{C}_1 ... \mathbf{C}_{\ppdegree}}$ is a fully
connected graph, and each 
partition $\mathbf{C}_j$ in the original communication graph $\mathbf{G}$ corresponds to a node in $\coG_{\mathbf{C}_1 ... \mathbf{C}_{\ppdegree}}$.

In the coarsened graph $\coG$, the weight on an edge between $\mathbf{C}_j$
and $\mathbf{C}_{j'}$ corresponds to the following --- \textit{if $\mathbf{C}_j$
and $\mathbf{C}_{j'}$ need to communicate
in a pipeline, what is the communicate cost 
of the optimal matching strategy between
devices in $\mathbf{C}_j$
and devices in $\mathbf{C}_{j'}$?}
Recall that $\ppbytes$ represents the amount of data
between two devices for pipeline parallel communication, 
we can model this cost by 

\vspace{-1.0em}
\begin{equation}
\begin{aligned} 
\min_{\mathcal{M}} \max_{(d, d') \in \mathcal{M}} 2 \left(\alpha_{d, d'} + \frac{\ppbytes}{\beta_{d,d'}} \right)
\end{aligned}
\label{eq:minmaxwpm}
\end{equation}
\vspace{-0.5em}

where $\mathcal{M}$ is a perfect matching
between $\mathbf{C}_j$ and $\mathbf{C}_{j'}$ ---
$(d, d') \in \mathcal{M}$ means that
device $d \in \mathbf{C}_j$ will communicate with device $d' \in \mathbf{C}_{j'}$ (i.e., they deal with the same micro-batch).
Computing this value is similar to
the classical minimal sum weight perfect matching problem (MinSumWPM) in 
bipartite graphs~\cite{goemans2009lecture}, with the only 
difference being that we compute the \textit{max} instead
of the \textit{sum}.  
As we will show in the supplementary material,
similar to MinSumWPM,
Eq~\ref{eq:minmaxwpm} can also be solved in 
PTIME.

The coarsened communication graph captures 
the pipeline parallel communication cost 
between two groups of devices, \textit{assuming}
they become neighbors in the pipeline.
Given this, we need to find an optimal permutation of $\mathbf{C}_1 ... \mathbf{C}_{\ppdegree}$, corresponds to the 
structure of the pipeline.
This becomes the \textit{open-loop traveling salesman problem}~\cite{papadimitriou1977euclidean}
 over this condensed graph (Figure~\ref{fig:1.1}(e)).
Formally, we have the following definition of the pipeline parallel cost: 

\vspace{-0.5em}
\begin{equation}
\begin{aligned} 
\textsc{PipelineP-Cost}\left(\mathbf{C}_1 ... \mathbf{C}_{\ppdegree}\right)
= \textsc{OpenLoopTSP}\left(\coG_{\mathbf{C}_1 ... \mathbf{C}_{\ppdegree}}\right)
\end{aligned}
\end{equation}
\vspace{-0.5em}

where $\coG_{\mathbf{C}_1 ... \mathbf{C}_{\ppdegree}}$ is the coarsened graph defined above.

\subsection{Searching via hybrid genetic algorithm}
\label{evol}

The scheduling problem solves the optimization problem in Eq~\ref{eq:main}, which 
corresponds to a \textit{balanced graph partition} problem with a complex objective corresponding to the communication cost.
Balanced graph partition problem is a challenging NP-hard problem~\cite{garey1979computers}. 
Over the years, researchers have been tackling this problem via different ways~\cite{andreev2006balanced,sanders2013think,bulucc2016recent}. 
We follow the line of research that uses hybrid genetic algorithm~\cite{el2006hybrid, kang2000hybrid} since it provides us the flexibility in dealing with complex objective.

{\bf Hybrid Genetic Algorithm.}
A hybrid genetic algorithm for balanced graph partition usually follows a structure as as follows. 
The input is a set of candidate balanced graph partitions which serves as the initial population. The algorithm 
generates the next 
generation as follows.
It first generates a new 
``offspring'' $o$ 
given two randomly selected
``parents'' $p_1$ and $p_2$.
One popular way is to 
randomly swap some 
nodes between these two 
parents (we follow \cite{kang2000hybrid}).
Given this offspring $o$,
we then conduct local search 
starting at $o$ to find 
a new balanced partitioning strategy $o^*$ that leads to 
better cost. 
We then add $o^*$ to the population and remove the worst partition candidate in the population if 
$o^*$ has a better cost.
As suggested by~\cite{el2006hybrid}, the combination of heuristic-based local search algorithms and genetic algorithm can accelerate convergence by striking the balance between local and global optimum.

{\bf Existing Local Search Strategy.}
The key in designing this algorithm is to come up with 
a good local search strategy.
For traditional graph partitioning task, one popular choice is to use the 
Kernighan-Lin Algorithm~\cite{bui1996genetic}.  
Which, at each iteration, tries to 
find a pair of nodes: $d$ in partition $\mathbf{C}_j$
and 
$d'$ in partition $\mathbf{C}_{j'}$,
to swap.
To find such a pair to swap,
it uses 
the following ``gain'' function:


\[
\begin{aligned} 
\textsc{gain}_{KL}( (d, \mathbf{C}_j) \leftrightarrow (d', \mathbf{C}_{j'})) = &\sum_{d'' \in \mathbf{C}_{j'}} 
    w_{d, d''}  -
\sum_{d'' \in \mathbf{C}_{j} - \left\{d\right\}} 
    w_{d, d''} \\
&+\sum_{d'' \in \mathbf{C}_{j}} 
    w_{d', d''}  -
\sum_{d'' \in \mathbf{C}_{j'} - \left\{d'\right\}} 
    w_{d', d''}
- 2 w_{d, d'}
\end{aligned}
\]

where $w_{i, j}$ corresponds to the
weight between node $i$
and $j$ in the graph.
However, directly applying this local search strategy, as we will also show in the experiment (Section~\ref{fig:evol}) does not work well. Greedily following
$\textsc{gain}_{KL}$ does not
decrease the communication cost of foundation model training. 
Therefore, we have to design a new 
local search strategy tailored 
to our cost model.

{\bf Improving Local Search Strategy.}
Our local search strategy is inspired by two observations:
\begin{enumerate}[leftmargin=2em,nosep,nolistsep]
\item Removing the device $d_1$
with a \textit{fast} connection (say with  $d_2$) within partition $\mathbf{C}_j$  will not tend to change the data parallel cost within 
$\mathbf{C}_j$, since it is only bounded
by the slowest connections.
\item Once $d_1$ is moved to 
 $\mathbf{C}_{j'}$, highly likely
 the pipeline parallel matching between 
 $\mathbf{C}_{j}$ and $\mathbf{C}_{j'}$ will consist of 
 the link $d_1 \leftrightarrow d_2$, since it is a fast connection.
\end{enumerate}

Therefore, in our local search strategy 
we only consider the fastest 
connection within 
$\mathbf{C}_j$: $d_1 \leftrightarrow d_2$
and the 
fastest 
connection within
$\mathbf{C}_{j'}$: $d_1' \leftrightarrow d_2'$
and generate four swap candidates: 
$d_1 \leftrightarrow d_1'$,
$d_1 \leftrightarrow d_2'$,
$d_2 \leftrightarrow d_1'$,
$d_2 \leftrightarrow d_2'$.
We use the following gain function (take $d_1 \leftrightarrow d_1'$ as an example): 

\[
\textsc{Gain}((d, \mathbf{C}_j) \leftrightarrow (d', \mathbf{C}_{j'})) = 
  \frac{1}{|\mathbf{C}_{j'}|} \sum_{d'' \in \mathbf{C}_{j'}} w_{d_1, d''} - 
w_{d_1, d_2} +
  \frac{1}{|\mathbf{C}_{j}|} \sum_{d'' \in \mathbf{C}_{j}} w_{d_1', d''} - 
w_{d_1', d_2'}
\]

where $\frac{1}{|\mathbf{C}_{j'}|} \sum_{d'' \in \mathbf{C}_{j'}} w_{d_1, d''}$ measures
the \textit{expected} pipeline 
parallel cost of connecting $d_1$ with
other devices in $\mathbf{C}_{j'}$ \textit{before the swap}, and $w_{d_1, d_2}$
is the cost of connecting $d_1$ with 
other devices in $\mathbf{C}_{j'}$ 
\textit{after the swap}, assuming this fast
link $d_1 \leftrightarrow d_2$
will now be used for pipeline parallelism.

Just like how Kernighan-Lin Algorithm~\cite{bui1996genetic}
can be extended to a circular version~\cite{kang2000hybrid} to swap multiple nodes beyond a pair, we can 
also extend our method into a circular one,
following procedure as circular KL with our
new gain function.

\subsection{Other System Optimizations}

We also have some system optimizations to further improve the performance. The most important optimization involves pipelining of communications and computations. We divide each stage in the pipeline into three slots: a receiving slot, a computation slot, and a sending slot. The receiving slot of stage $j$ needs to build connections to receive activations from the stage $j-1$ in forward propagation and to receive gradients of activations from stage $j+1$. The computation slot handles the computation in forward and backward propagation. Symmetric to the receiving slot, the sending slot of stage $j$ needs to build connections to send activations to stage $j+1$  in the forward propagation and send gradients of activations to stage $j-1$ in the backward propagations. These three slots are assigned to three CUDA streams so that they will be further pipelined efficiently; as a result,  communication will overlap with computation.
In the decentralized scenario (communication bound),  computation can be fully hidden inside the communication time.   

\section{Evaluation}
\label{sec:exp}

We demonstrate that our system can speed up foundation model training in decentralized setting. Specifically, (1) We show that our system is 4.8$\times$ faster, in end-to-end runtime, than the state-of-the-art systems (Megatron and Deepspeed) training GPT3-1.3B in world-wide geo-distributed setting. Surprisingly, it is only $1.7-2.3\times$ slower than these systems in data centers. This indicates that we can bridge the gap between decentralized and data center training (up to $100\times$ slower networks) through scheduling and system optimization; (2) We demonstrate the necessity of our scheduler through an ablation study. We show that with the scheduler, our system is 2.7$\times$ faster in  world-wide geo-distributed setting. 

\begin{figure}[t!]
    \centering
    \includegraphics[width=0.975\linewidth]{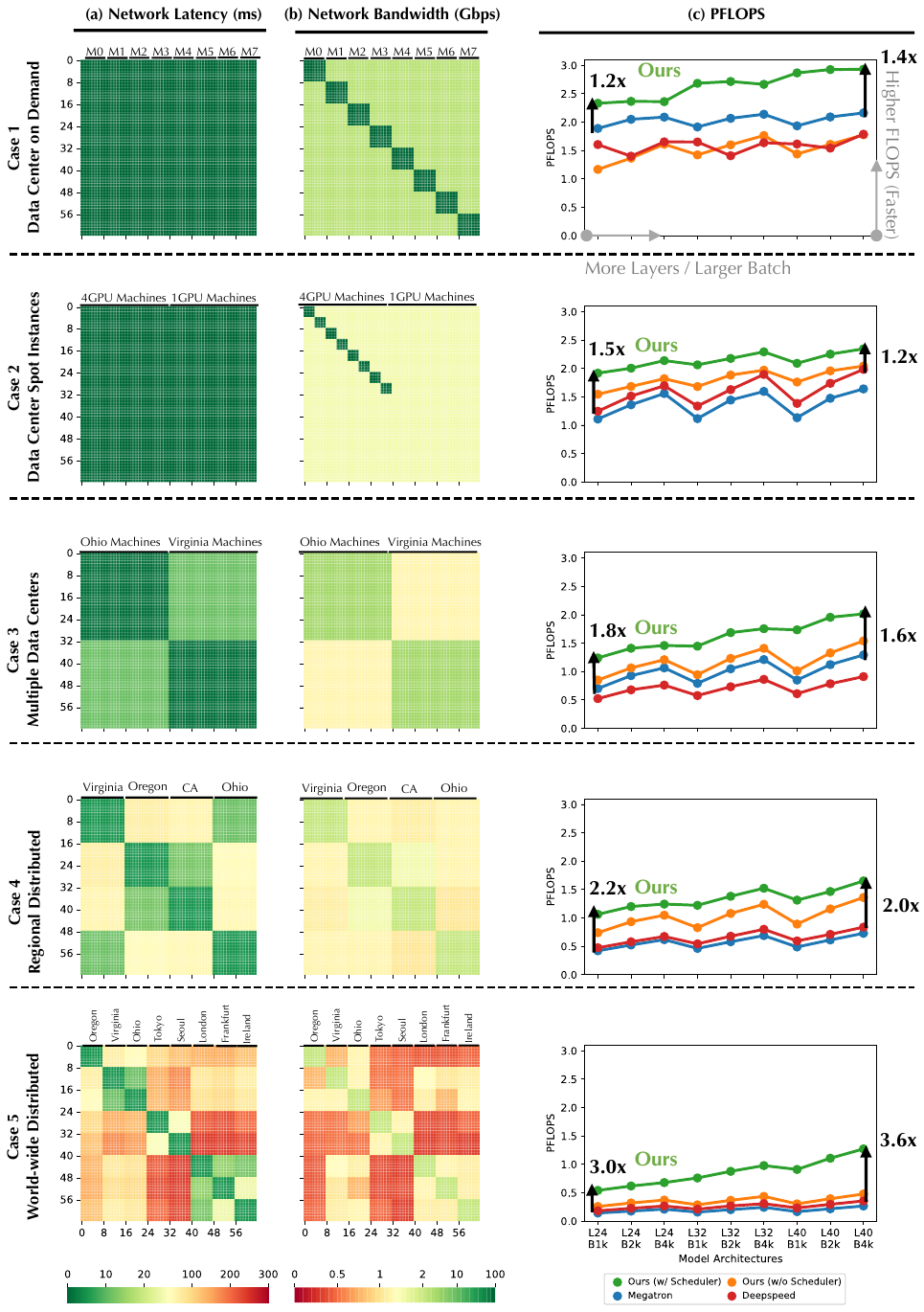}
    \vspace{-0.5em}
    \caption{End to end compassion of our system with Megatron and Deepspeed in 5 different scenarios. Column (a) and (b) visualize the delay and bandwidth of 5 scenario respectively; Column (c) illustrate the comparison of Megatron, Deepspeed and our system w and w/o scheduler.}
    \label{fig:exp}
    \vspace{-1.5em}
\end{figure}

\subsection{Experimental Setup.} 

To simulate the decentralized 
setting, we use 8 different 
AWS regions (Oregon, Virginia, Ohio, Tokyo, Seoul, London, Frankfurt, and Ireland)
and measure the latency and bandwidth between these regions
(we consider the bandwidth that
we can realistically obtain
using NCCL and UDP 
hole punching between 
these regions). Given these
measurements, we use 64 Tesla V100 GPUs and control their pairwise
communication latency and bandwidth
for five different cases:

\emph{\underline{Case 1. Data center on demand.}} This is a standard setting that a user can obtain to train foundation models. we use 8 AWS \texttt{p3.16xlarge} nodes (each with 8 V100 GPUs); the intra-node connection is NVLink of 300 GB/s bi-directional bandwidth (150 GB/s unidirectional), and the inter-node connection has a bandwidth of 25 Gbps. We do not manually control latency and bandwidth here.

\emph{\underline{Case 2. Data center spot instances.}} Spot GPUs are cheaper in a data center, but can be located on different types of machine. In this case, we rent 4 AWS \texttt{p3.8xlarge} nodes (each with 4 V100) and 32 \texttt{p3.2xlarge} nodes (each with 1 V100); the intra- \texttt{p3.8xlarge} node connection has a bandwidth of 100 Gbps, and the inter-node connection has a bandwidth of 10 Gbps. We do not manually control latency and bandwidth in this case.

\emph{\underline{Case 3. Multiple Data Centers.}} We consider two organizations, one in Ohio and another in Virginia, each organization contributes 32 V100 GPUs; within each organization, the bandwidth is 10 Gbps, and connections cross different campuses have a delay of 10 ms and bandwidth of 1.12 Gbps.

\emph{\underline{Case 4. Regional geo-distributed.}} We consider individual GPUs cross four different regions in US (California, Ohio, Oregon, and Virginia) ; within each region, the delay is 5 ms and bandwidth is 2 Gbps; cross different regions, the delay is $10{\sim}70$ms and the bandwidth is  $1.0{\sim}1.3$ Gbps. 

\emph{\underline{Case 5. World-wide geo-distributed.}} We consider individual GPUs cross eight different regions world-wide (Oregon, Virginia, Ohio, Tokyo, Seoul, London, Frankfurt, and Ireland); within each region, the delay is 5 ms and bandwidth is 2 Gbps; cross different regions, the delay is $10{\sim}250$ms and the bandwidth is $0.3{\sim}1.3$Gbps.

{\bf Metrics and Model Architecture.}
Since we do not introduce any 
optimizations that might change 
the computation or convergence,
we can compare all methods 
by its throughput, we can compare all systems by the total number of 
floating point
operations per second (PFLOPS),
which is inverse proportional 
to the \textit{runtime of each iteration}
(which we show in Appendix). We use the 
standard GPT3-1.3B architecture~\cite{brown2020language}, while also benchmarked different number of layers $\{24, 32, 40\}$, and batch sizes $\{1024, 2048, 4096\}$.

{\bf Tuning of Megatron and Deepspeed.}
We did a careful grid search of different parallelism settings and report the optimal results in each case---in Case 1, the optimal setting includes tensor model parallelism in Megatron and ZeRO-S3 in Deepspeed; in all other cases, the optimal settings are based on pipeline and data parallelism. We discuss more details in Appendix. 

\subsection{Results}

\paragraph*{End-to-end Comparison.} 

Figure \ref{fig:exp}(c) shows the end-to-end comparison in terms of 
averaged PFLOPS achieved
across different settings and different batch sizes and number of layers. 
In the world-wide geo-distributed cases, we achieve an $4.8\times$ speedup of Megatron, ($3.6\times$ speedup of Deepspeed). While in all other cases, our system can be  $1.2-2.5\times$ faster. 
If we compare our system in Case 5 (world-wide geo-distributed)
and Megatron/Deepspeed in Case 1 (data center on demand), it is exciting to see that the performance slowdown caused by decentralization 
is only $1.7-3.5\times$! This 
illustrates the great potential 
of decentralized training for foundation models.
Additionally, Figure \ref{fig:exp}(c) illustrates another interesting behavior pattern. As increasing the batch size does not 
increases the communication cost of
data parallelism 
and 
increasing \# layers per device does not 
increases the communication cost
of pipeline parallelism, 
with a larger batch size and a deeper model, the gap between 
centralized Megatron/Deepspeed and our decentralized system is even smaller.


\begin{figure}[t!]
    \centering
    \includegraphics[width=0.975\linewidth]{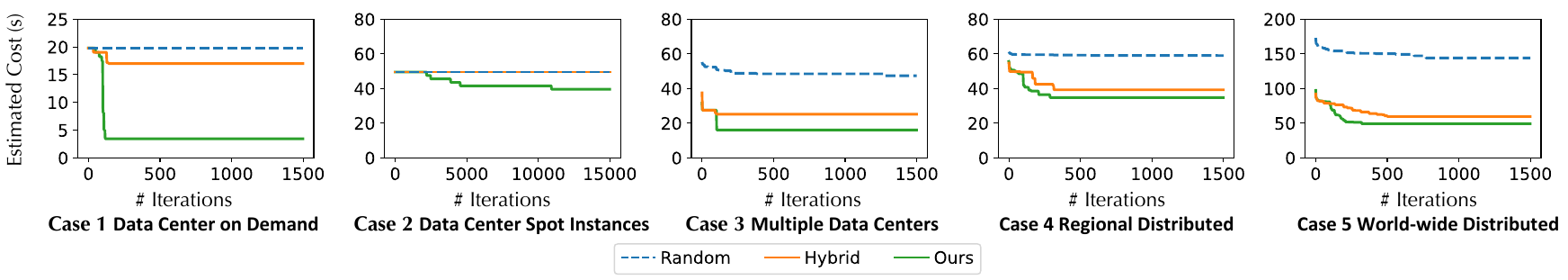}
    \vspace{-0.5em}
    \caption{Comparison of Search Strategies.}
    \label{fig:evol}
    \vspace{-0.5em}
\end{figure}

\paragraph*{Effectiveness of Scheduler.}

To evaluate the effectiveness of 
the scheduler, we disable it
and use a random assignment
in all cases and the results 
are also illustrated in 
Figure~\ref{fig:exp}(c).
We see that with 
our scheduler provides up to
$2.7\times$ speeds up.
To evaluate 
our local search strategy, we also compare our scheduler with
a scheduler that uses 
the standard  
Kernighan-Lin algorithm for local search, illustrated in Figure \ref{fig:evol}.
We see that, while both outperform random, our carefully designed 
local search strategy significantly
outperforms Kernighan-Lin.

\section{Related Work}

\textbf{Foundation models.} Foundation models\cite{bommasani2021opportunities} refer to models that are trained on large-scale data and can be adapted (e.g., fine-tuned) to a wide range of downstream tasks. Current examples include BERT~\cite{devlin2018bert}, GPT-3~\cite{brown2020language}, and CLIP\cite{radford2021learning}. Foundation models are usually trained in a data center, where the connection between GPUs is fast and homogeneous. ML infrastructures such as Megatron\cite{shoeybi2019megatron} and ZeRO\cite{rajbhandari2020zero,ren2021zero} have been proposed to distribute the training of these foundation models in a data center. Megatron uses \texttt{AllReduce} to synchronize activations in tensor model parallelism; ZeRO adopts \texttt{ScatterGather} to dispatch sharded parameters for layer-wise data parallelism. However, such collective communication paradigms would cause serious performance problems with slow and heterogeneous connections (see Appendix for detailed discussions).

\vspace{-0.5em}
\textbf{Decentralized optimization}. Decentralized training is first proposed within the scope of data parallelism, where each worker only synchronizes gradients with its neighbors (instead of all workers) to remove the latency bottleneck~\cite{koloskova2019decentralized,li2018pipe,lian2017can,lian2018asynchronous,tang2018communication,tang2018d}. Recently, \cite{diskin2021distributed,ryabinin2021swarm} has also modified the implementation of data or pipeline parallelism to support training in an open collaborative environment. Varuna~\cite{athlur2022varuna} is released by Microsoft to support the training of GPT models in spot instances from a cloud service provider, which has the potential to be extended to the open collective scenario, but there is limited consideration with respect to the challenges of heterogeneous connections.   


\vspace{-0.5em}
\textbf{Volunteer computing}. Distributing computationally intensive tasks over an open collaborative environment has been advocated for a few decades since the development of BOINC\cite{anderson2004boinc}; for example, the folding@home project~\cite{shirts2000screen} has been running simulations about protein dynamics on volunteers' personal computers for more than 20 years.
Recently, the learning@home project\cite{ryabinin2020towards} starts to consider training of mixture-of-expert transformers in such a volunteer computing setting.

\section{Conclusion}
In this paper, we probe the opportunity to train foundation models via a decentralized training regime with devices connected over a heterogeneous network.
We propose an effective scheduling algorithm to assign tasklets from the foundation model pre-train computation. Empirical studies suggest that, in the worldwide geo-distributed scenario, our proposed scheduling algorithm enables a $4.8\times$ speed-up compared to prior state-of-the-art training systems.  We believe that the decentralization and democratization of the training of FMs can shift the balance of power positively, but also necessitate new governance structures to help ensure the responsible development and deployment of FMs.

\section*{Acknowledgments}
\begin{footnotesize}
CZ and the DS3Lab gratefully acknowledge the support from the Swiss State Secretariat for Education, Research and Innovation (SERI) under contract number MB22.00036 (for European Research Council (ERC) Starting Grant TRIDENT 101042665), the Swiss National Science Foundation (Project Number 200021\_184628, and 197485), Innosuisse/SNF BRIDGE Discovery (Project Number 40B2-0\_187132), European Union Horizon 2020 Research and Innovation Programme (DAPHNE, 957407), Botnar Research Centre for Child Health, Swiss Data Science Center, Alibaba, Cisco, eBay, Google Focused Research Awards, Kuaishou Inc., Oracle Labs, Zurich Insurance, and the Department of Computer Science at ETH Zurich. CR gratefully acknowledges the support of NIH under No. U54EB020405 (Mobilize), NSF under Nos. CCF1763315 (Beyond Sparsity), CCF1563078 (Volume to Velocity), and 1937301 (RTML); ARL under No. W911NF-21-2-0251 (Interactive Human-AI Teaming); ONR under No. N000141712266 (Unifying Weak Supervision); ONR N00014-20-1-2480: Understanding and Applying Non-Euclidean Geometry in Machine Learning; N000142012275 (NEPTUNE); NXP, Xilinx, LETI-CEA, Intel, IBM, Microsoft, NEC, Toshiba, TSMC, ARM, Hitachi, BASF, Accenture, Ericsson, Qualcomm, Analog Devices, Google Cloud, Salesforce, Total, the HAI-GCP Cloud Credits for Research program, the Stanford Data Science Initiative (SDSI), and members of the Stanford DAWN project: Facebook, Google, and VMWare. The U.S. Government is authorized to reproduce and distribute reprints for Governmental purposes notwithstanding any copyright notation thereon. Any opinions, findings, and conclusions or recommendations expressed in this material are those of the authors and do not necessarily reflect the views, policies, or endorsements, either expressed or implied, of NIH, ONR, or the U.S. This work was supported by an Open Philanthropy Award. The computation required in this work was provided by Together Computer \url{https://together.xyz/}.
\end{footnotesize}

\bibliographystyle{unsrtnat}
\bibliography{main}

\clearpage

\section{Social Impact of Decentralized Training}
In this paper, we find that decentralized training shows great potential for foundation models --- such a technique would lead to significant positive social impacts. For example, decentralized training can utilize more inexpensive computational resources, which can significantly reduce the budget for the training of foundation models. This would increase the accessibility of foundation models for small research and commercial institutions. In fact, if the expense would be significantly reduced, large organizations would also receive some benefit from adopting such technology. On the other hand, we also notice that decentralization and democratization can also lead to a lack of control of cheaper computing resources and accelerate the risks of foundation models~\cite{bommasani2021opportunities}. We look forward to actively engaging the community on governance questions.

\section{Limitation and Future Work}

There are still some limitations of the current approach that could be explored further in future work.

First, we assume a homogeneous compute GPU resource in the scheduling algorithm, and in practice, different types of GPU could join the training computations. A native extension of the current solution could be to further split the tasklets into smaller pieces and to assign different numbers of pieces in different types of GPUs considering their memory budget and compute power as constraints. However, there are many opportunities for further improvement.

Second, there is still lots of room for the improvement of our scheduling algorithm, and strengthen our argument about the end-to-end speedup. In this paper, we hope to first provide some positive insights to the community that the potential of decentralized training can make it deployable for giant foundation models. On the other hand, there are many recent advances in system optimization that could lead to further improvement. For example, exploring recent advances in reinforcement learning~\cite{shyalika2020reinforcement} to solve our scheduling algorithm would be an interesting future direction. We believe that any improvement there can only improve the decentralized performance, which is consistent with the central message we try to share in this paper.

Last but not least, there are still some important open questions on the system side to handle the dynamics in decentralized environments. For example, some mechanism should be necessary to handle the dynamic join and leave of GPU nodes. On the other hand, failure happens more frequently in the decentralized environment, as fault tolerance should be considered for deployment, we believe that the current strategy such as checkpointing the model periodically could be adopted for this problem, but there would be some suitable solutions for the decentralized training runtime.

\section{Anatomy of the Current ML Systems for Foundation Model Training}
\label{sec:anatomy}

Training foundation models~\cite{bommasani2021opportunities} is a challenging task due to the enormous scale of such models --- even the most powerful GPU cannot hold a complete copy of parameters for such models \cite{smith2022using}. Thus, one cannot train such a model without distribution or using vanilla data parallelism. 


Two popular approaches have been proposed to distribute the training of such foundation models in a data center: 
\begin{itemize}[leftmargin=*,nosep,nolistsep]
\item Megatron~\cite{shoeybi2019megatron} distributes training by combining its proposed tensor model parallelism with pipeline parallelism~\cite{huang2019gpipe, narayanan2019pipedream} and data parallelism~\cite{li2020pytorch}. The tensor model parallelism partitions individual layers across a group of workers and must run one \texttt{AllReduce} for the output activations of \textit{each} layer in forward propagation and one \texttt{AllReduce} for the corresponding gradients in backward propagation for \textit{each} layer.      

\item ZeRO~\cite{rajbhandari2020zero} can be viewed as an effective optimization for data parallelism. The most effective mode is called ZeRO stage-3 from the Deepspeed implementation~\cite{rasley2020deepspeed}, and the equivalent implementation is known as Fully Sharded Data Parallelism (FSDP) from Fairscale~\cite{baines2021fairscale}). In this mode, the parameter is sharded among all workers --- in forward propagation, each worker conducts one \texttt{AllGather} to collect the parameters demanded for the current layer and discard the parameter after the forward computation; in backward propagation, each worker uses one \texttt{AllGather} to collect the parameter again and run one 
\texttt{ReduceScatter} to synchronize the gradients of this layer after the backward computation. 

\end{itemize}


Both Megatron and ZeRO take a heavy usage of collective communication paradigms~\cite{chan2007collective}, which leads to two fundamental problems when it is deployed with \textit{heterogeneous} and \textit{slow} connections:
\begin{itemize}[leftmargin=*,nosep,nolistsep]
\item \textit{Demanding high bandwidth connections.} Both Megatron and ZeRO require high bandwidth connections for collective communications, since the compute cores are \textit{idled} during communication slots. As long as communication takes an increasing share (due to lower bandwidth) of the execution time, the hardware efficiency drops dramatically. In fact, tensor model parallelism is recommended only within a single DGX server equipped with high-bandwidth NVLinks~\cite{smith2022using}.     

\item \textit{Sensitive to straggler.} The design and implementation of state-of-the-art collective communication libraries, e.g., NCCL~\cite{nccl}, assume highly homogeneous connections within a data center, thus there is not sufficient robustness to handle the straggler among workers due to the heterogeneity of the open collective runtime. Furthermore, the layer-wise usage of collective communications in both Megatron and ZeRO has intensified this problem. 
\end{itemize} 


To bridge the performance gap between the data center and the decentralized open environment, we need to rethink the communication paradigms in different parallel strategies. 

\begin{itemize}[leftmargin=*,nosep,nolistsep]
\item \textit{Pipeline parallelism is communication efficient.} Pipeline parallelism~\cite{huang2019gpipe, narayanan2019pipedream, narayanan2021memory} partitions the model into multiple stages and a batch into multiple mini-batches, where once a worker finished the forward computation of a micro-batch, this worker will send the activations to the worker running the next stages; on the other hand, a worker needs to send the gradients of the activation back to the last stage in the backward propagation. Notice that pipeline parallelism utilizes \textit{point-to-point} communications instead of collective paradigms. As long as one can put an increasing amount of computation inside a stage, the ratio of communication cost will also drop, leading to more efficient utilization of compute cores.\footnote{Notice this is not always true since the device memory is limited. However, one can offload~\cite{ren2021zero} (e.g., activations and parameters) to host memory to perform training on larger models with limited GPU device memory. Furthermore, the offloading through PCI-e is much faster compared to the decentralized connections, although it is slower than NVLink between GPUs in a data center.} On the other hand, pipeline parallelism has its own limitation --- one can only partition a model to a limited number of stages, which cannot scale out to lots of GPUs. We need to combine pipeline parallelism with data parallelism to scale out the training.

\item \textit{Scheduling is essential.} 
The \textit{point-to-point} communication pattern in pipeline parallelism provides good opportunities to assign the training procedure on the decentralized environment that utilizes fast links and avoids slow links by a carefully designed scheduler, as presented in Section \ref{sec:schedule}. 
\end{itemize}

\section{Additional Details of Experimental Evaluation}
We enumerate some additional details about our experiments. 

\subsection{Multiple Execution of the Benchmark}

We repeated all the benchmarks of 5 different scenarios listed in Section \ref{sec:exp} three times. For our system with scheduler, since the scheduled layout is the same, we simply issued three independent executions in each scenarios; For our system without scheduler, we used three different random seeds (2022, 2023, 2024) to generate three layouts, and issued one execution for each layout in each scenario. The number in Figure \ref{fig:exp} is based on an average of three different runs for each scenario --- to avoid visual confusion, we did not plot the error bar within this line plot. We also repeated the scheduling algorithms three times with random seeds (0, 1, 2) to generate scheduled layouts and reported the average estimated cost (seconds) in Figure \ref{fig:evol}.
In Figure \ref{fig:exp_appendix}, we plot \textit{runtime of each iteration} as a bar chart with error bars. Notice that the variance of all executions in each setting is within $5\%$.

\subsection{Tuning of Megatron and Deepspeed}

\textbf{Megatron.} We carefully tuned Megatron to perform a fair comparison with our system. As we mentioned in Section \ref{sec:anatomy}. Megatron has three free degrees of parallel strategies: \textit{tensor model parallelism}, \textit{pipeline parallelism}, and \textit{data parallelism}, we note the degrees of these parallel strategies as $\tpdegree$, $\ppdegree$, and $\dpdegree$ respectively. We conduct a complete grid search of the combinations of these hyper-parameters in the space of:  
\[
\left\{\left(\tpdegree, \ppdegree, \dpdegree\right) |  \tpdegree, \ppdegree, \dpdegree\in \{1, 2, 4, 8\} \text{ and } \tpdegree \times \ppdegree \times \dpdegree=64 \right\}.
\] 
And we reported the optimal setting as the results for Megatron. Interestingly, only in Case 1 (data center on demand), the optimal setting includes tensor model parallelism (i.e., $\tpdegree \neq 1$), where $\tpdegree=2, \ppdegree=4, \dpdegree=8$; in all other scenarios, the optimal setting is $\tpdegree=1, \ppdegree=8, \dpdegree=8$. This illustrates that tensor model parallelism is not suitable in slow and heterogeneous settings, consistent with our analysis in Section \ref{sec:anatomy}.  Since Megatron does not include a similar scheduler as its own components, we use the same random layouts as what we test for our system without scheduler.  

\textbf{Deepspeed.} To run Deepspeed, we start with ZeRO-S3, which is usually viewed as the most significant technical contribution of the Deepspeed system. Under the same settings, the execution time of ZeRO-S3 is much longer comparing to both Megatron and our system (See Figure \ref{fig:exp_zeros3}). This is consistent with our analysis in Appendix B. We then try to combine the pipeline parallel implementation in Deepspeed with its different implementations of data parallelism (e.g., ZeRO-S 1, 2 and 3)---it turns out that even the latest version of Deepspeed (0.6.7) only allows ZeRO-S1 to combine with pipeline parallelism. We find this combination outperforms ZeRO-S3 in most of the settings in Case 1 and all settings in Case 2,3,4 and 5. Notice that results of Deepspeed we report in Figure \ref{fig:exp} is based on the optimal result of these two settings.

\begin{figure}[t!]
    \centering
    \includegraphics[width=0.95\linewidth]{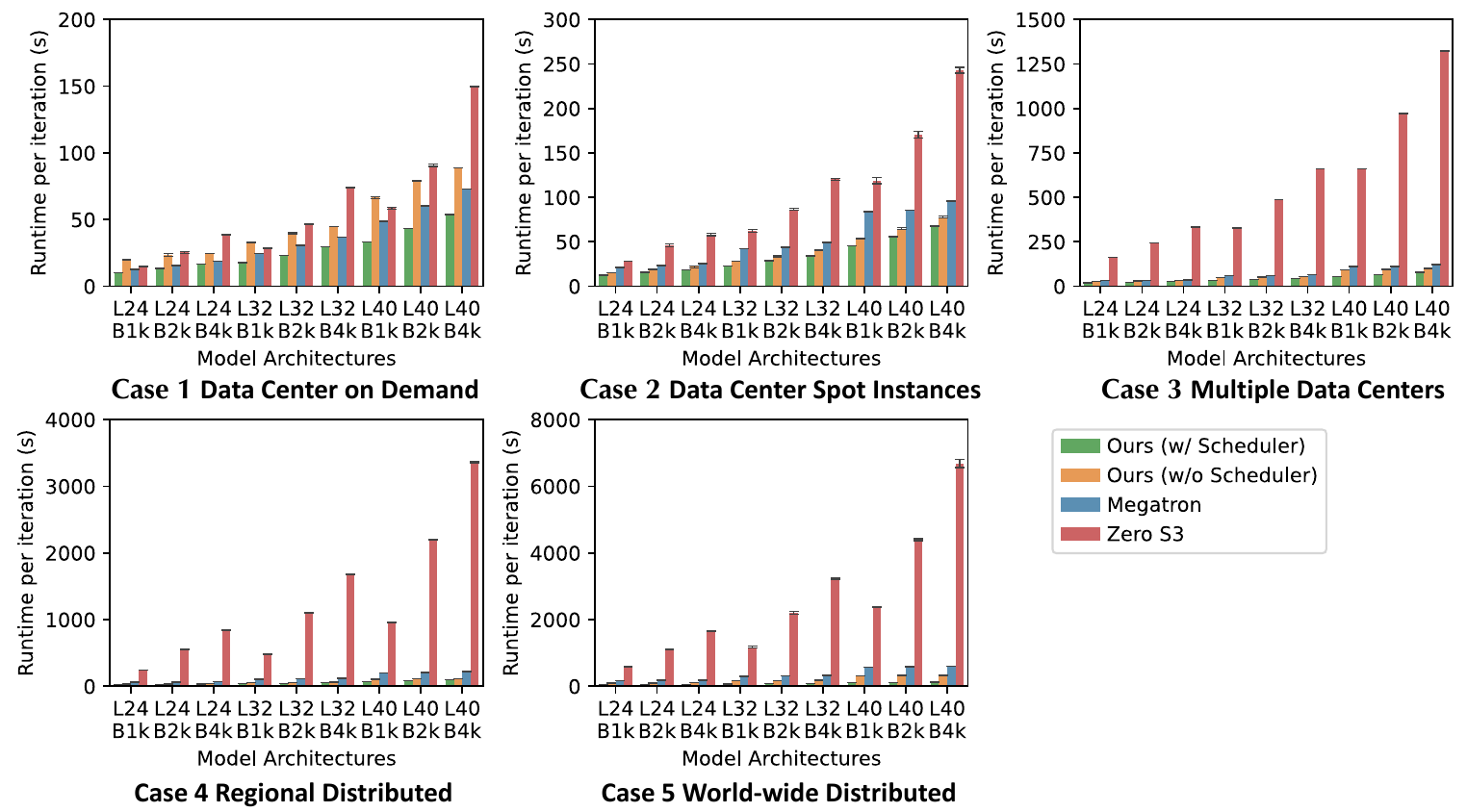}
    \caption{The performance of ZeRO-S3 from Deepspeed in terms of \textit{runtime of each iteration} in 5 different Scenarios.}
    \label{fig:exp_zeros3}
\end{figure}

\subsection{Network Benchmark.}

To obtain the network delay and bandwidth between different regions across the world, we rent AWS instances in 9 different data centers (California, Oregon, Virginia, Ohio, Tokyo, Seoul, London, Frankfurt, and Ireland). Instead of using AWS VPC, we setup our own VPN (using StrongSwan~\cite{strongswan}) established on the public IP of these instances---any GPU machine connected to Internet can be linked in the same way. The strongSwan VPN would expose a private IP associated with a visible network interface, we can bind the NCCL communication on this network interface. The delay and bandwidth we obtained for cross-region NCCL connections are summarized in Table \ref{tab:net4} for Case 4 regional geo-distributed scenario and and Table \ref{tab:net5} for Case 5 world-wide geo-distributed scenario.

\begin{table}[t!]
    \centering
{
\begin{tabular}{|lllll|}
\hline
\multicolumn{5}{|c|}{Delay (ms)}                                                                                                             \\ \hline
\multicolumn{1}{|l|}{}           & \multicolumn{1}{l|}{California} & \multicolumn{1}{l|}{Ohio} & \multicolumn{1}{l|}{Oregon} & Virginia \\ \hline
\multicolumn{1}{|l|}{California} & \multicolumn{1}{l|}{-}           & \multicolumn{1}{l|}{52}     & \multicolumn{1}{l|}{12}       & \multicolumn{1}{l|}{59} \\ \hline
\multicolumn{1}{|l|}{Ohio}       & \multicolumn{1}{l|}{52}           & \multicolumn{1}{l|}{-}     & \multicolumn{1}{l|}{49}       & \multicolumn{1}{l|}{11} \\ \hline
\multicolumn{1}{|l|}{Oregon}     & \multicolumn{1}{l|}{12}           & \multicolumn{1}{l|}{49}     & \multicolumn{1}{l|}{-}       & \multicolumn{1}{l|}{67} \\ \hline
\multicolumn{1}{|l|}{Virginia}   & \multicolumn{1}{l|}{59}           & \multicolumn{1}{l|}{11}     & \multicolumn{1}{l|}{67}       & \multicolumn{1}{l|}{-} \\ \hline
\multicolumn{5}{|c|}{Bandwidth (Gbps)}                                                                                                         \\ \hline
\multicolumn{1}{|l|}{}           & \multicolumn{1}{l|}{California} & \multicolumn{1}{l|}{Ohio} & \multicolumn{1}{l|}{Oregon} & Virginia \\ \hline
\multicolumn{1}{|l|}{California} & \multicolumn{1}{l|}{-}           & \multicolumn{1}{l|}{1.02}     & \multicolumn{1}{l|}{1.25}       & \multicolumn{1}{l|}{1.05} \\ \hline
\multicolumn{1}{|l|}{Ohio}       & \multicolumn{1}{l|}{1.02}           & \multicolumn{1}{l|}{-}     & \multicolumn{1}{l|}{1.10}       & \multicolumn{1}{l|}{1.12} \\ \hline
\multicolumn{1}{|l|}{Oregon}     & \multicolumn{1}{l|}{1.25}           & \multicolumn{1}{l|}{1.10}     & \multicolumn{1}{l|}{-}       & \multicolumn{1}{l|}{1.15} \\ \hline
\multicolumn{1}{|l|}{Virginia}   & \multicolumn{1}{l|}{1.05}           & \multicolumn{1}{l|}{1.12}     & \multicolumn{1}{l|}{1.15}       & \multicolumn{1}{l|}{-}\\ \hline
\end{tabular}}
\vspace{1em}
\caption{Delay (in milliseconds) and bandwidth (in Gbps) obtained by NCCL for Case 4 regional distributed scenario.}
\label{tab:net4}
\end{table}

\begin{table}[t!]
    \centering
{
\begin{tabular}{|lllllllll|}
\hline
\multicolumn{9}{|c|}{Delay (ms)}                                                                                                                                                                                                                              \\ \hline
\multicolumn{1}{|l|}{}          & \multicolumn{1}{l|}{Oregon} & \multicolumn{1}{l|}{Virginia} & \multicolumn{1}{l|}{Ohio}  & \multicolumn{1}{l|}{Tokyo} & \multicolumn{1}{l|}{Seoul} & \multicolumn{1}{l|}{London} & \multicolumn{1}{l|}{Frankfurt} & Ireland \\ \hline
\multicolumn{1}{|l|}{Oregon}    & \multicolumn{1}{l|}{-}      & \multicolumn{1}{l|}{67}       & \multicolumn{1}{l|}{49}    & \multicolumn{1}{l|}{96}    & \multicolumn{1}{l|}{124}   & \multicolumn{1}{l|}{136}    & \multicolumn{1}{l|}{143}       & 124     \\ \hline
\multicolumn{1}{|l|}{Virginia}  & \multicolumn{1}{l|}{67}     & \multicolumn{1}{l|}{-}        & \multicolumn{1}{l|}{11}    & \multicolumn{1}{l|}{143}   & \multicolumn{1}{l|}{172}   & \multicolumn{1}{l|}{76}     & \multicolumn{1}{l|}{90}        & 67      \\ \hline
\multicolumn{1}{|l|}{Ohio}      & \multicolumn{1}{l|}{49}     & \multicolumn{1}{l|}{11}       & \multicolumn{1}{l|}{-}     & \multicolumn{1}{l|}{130}   & \multicolumn{1}{l|}{159}   & \multicolumn{1}{l|}{86}     & \multicolumn{1}{l|}{99}        & 77      \\ \hline
\multicolumn{1}{|l|}{Tokyo}     & \multicolumn{1}{l|}{96}     & \multicolumn{1}{l|}{143}      & \multicolumn{1}{l|}{130}   & \multicolumn{1}{l|}{-}     & \multicolumn{1}{l|}{34}    & \multicolumn{1}{l|}{210}    & \multicolumn{1}{l|}{235}       & 199     \\ \hline
\multicolumn{1}{|l|}{Seoul}     & \multicolumn{1}{l|}{124}    & \multicolumn{1}{l|}{172}      & \multicolumn{1}{l|}{159}   & \multicolumn{1}{l|}{34}    & \multicolumn{1}{l|}{-}     & \multicolumn{1}{l|}{238}    & \multicolumn{1}{l|}{235}       & 228     \\ \hline
\multicolumn{1}{|l|}{London}    & \multicolumn{1}{l|}{136}    & \multicolumn{1}{l|}{76}       & \multicolumn{1}{l|}{86}    & \multicolumn{1}{l|}{210}   & \multicolumn{1}{l|}{238}   & \multicolumn{1}{l|}{-}      & \multicolumn{1}{l|}{14}        & 12      \\ \hline
\multicolumn{1}{|l|}{Frankfurt} & \multicolumn{1}{l|}{143}    & \multicolumn{1}{l|}{90}       & \multicolumn{1}{l|}{99}    & \multicolumn{1}{l|}{235}   & \multicolumn{1}{l|}{235}   & \multicolumn{1}{l|}{14}     & \multicolumn{1}{l|}{-}         & 24      \\ \hline
\multicolumn{1}{|l|}{Ireland}   & \multicolumn{1}{l|}{124}    & \multicolumn{1}{l|}{67}       & \multicolumn{1}{l|}{77}    & \multicolumn{1}{l|}{199}   & \multicolumn{1}{l|}{228}   & \multicolumn{1}{l|}{12}     & \multicolumn{1}{l|}{24}        & -       \\ \hline
\multicolumn{9}{|c|}{Bandwidth (Gbps)}                                                                                                                                                                                                                        \\ \hline
\multicolumn{1}{|l|}{}          & \multicolumn{1}{l|}{Oregon} & \multicolumn{1}{l|}{Virginia} & \multicolumn{1}{l|}{Ohio}  & \multicolumn{1}{l|}{Tokyo} & \multicolumn{1}{l|}{Seoul} & \multicolumn{1}{l|}{London} & \multicolumn{1}{l|}{Frankfurt} & Ireland \\ \hline
\multicolumn{1}{|l|}{Oregon}    & \multicolumn{1}{l|}{-}      & \multicolumn{1}{l|}{1.15}     & \multicolumn{1}{l|}{1.10}  & \multicolumn{1}{l|}{0.523} & \multicolumn{1}{l|}{0.46}  & \multicolumn{1}{l|}{0.42}   & \multicolumn{1}{l|}{0.404}     & 0.482   \\ \hline
\multicolumn{1}{|l|}{Virginia}  & \multicolumn{1}{l|}{1.15}   & \multicolumn{1}{l|}{-}        & \multicolumn{1}{l|}{1.12}  & \multicolumn{1}{l|}{0.524} & \multicolumn{1}{l|}{0.500} & \multicolumn{1}{l|}{0.364}  & \multicolumn{1}{l|}{1.02}      & 1.05    \\ \hline
\multicolumn{1}{|l|}{Ohio}      & \multicolumn{1}{l|}{1.10}   & \multicolumn{1}{l|}{1.12}     & \multicolumn{1}{l|}{-}     & \multicolumn{1}{l|}{0.694} & \multicolumn{1}{l|}{0.529} & \multicolumn{1}{l|}{1.05}   & \multicolumn{1}{l|}{0.799}     & 1.14    \\ \hline
\multicolumn{1}{|l|}{Tokyo}     & \multicolumn{1}{l|}{0.523}  & \multicolumn{1}{l|}{0.524}    & \multicolumn{1}{l|}{0.694} & \multicolumn{1}{l|}{-}     & \multicolumn{1}{l|}{1.1}   & \multicolumn{1}{l|}{0.366}  & \multicolumn{1}{l|}{0.36}      & 0.465   \\ \hline
\multicolumn{1}{|l|}{Seoul}     & \multicolumn{1}{l|}{0.46}   & \multicolumn{1}{l|}{0.500}    & \multicolumn{1}{l|}{0.529} & \multicolumn{1}{l|}{1.1}   & \multicolumn{1}{l|}{-}     & \multicolumn{1}{l|}{0.342}  & \multicolumn{1}{l|}{0.358}     & 0.335   \\ \hline
\multicolumn{1}{|l|}{London}    & \multicolumn{1}{l|}{0.42}   & \multicolumn{1}{l|}{0.364}    & \multicolumn{1}{l|}{1.05}  & \multicolumn{1}{l|}{0.366} & \multicolumn{1}{l|}{0.342} & \multicolumn{1}{l|}{-}      & \multicolumn{1}{l|}{1.14}      & 1.09    \\ \hline
\multicolumn{1}{|l|}{Frankfurt} & \multicolumn{1}{l|}{0.404}  & \multicolumn{1}{l|}{1.02}     & \multicolumn{1}{l|}{0.799} & \multicolumn{1}{l|}{0.36}  & \multicolumn{1}{l|}{0.358} & \multicolumn{1}{l|}{1.14}   & \multicolumn{1}{l|}{-}         & 1.08    \\ \hline
\multicolumn{1}{|l|}{Ireland}   & \multicolumn{1}{l|}{0.482}  & \multicolumn{1}{l|}{1.05}     & \multicolumn{1}{l|}{1.14}  & \multicolumn{1}{l|}{0.465} & \multicolumn{1}{l|}{0.335} & \multicolumn{1}{l|}{1.09}   & \multicolumn{1}{l|}{1.08}      & -       \\ \hline
\end{tabular}}
\vspace{1em}
\caption{Delay (in milliseconds) and bandwidth (in Gbps) obtained by NCCL for Case 5 world-wide distributed scenario.}
\label{tab:net5}
\end{table}

\subsection{Other Presentation of Experimental Results}

\begin{figure}[t!]
    \centering
    \includegraphics[width=0.9\linewidth]{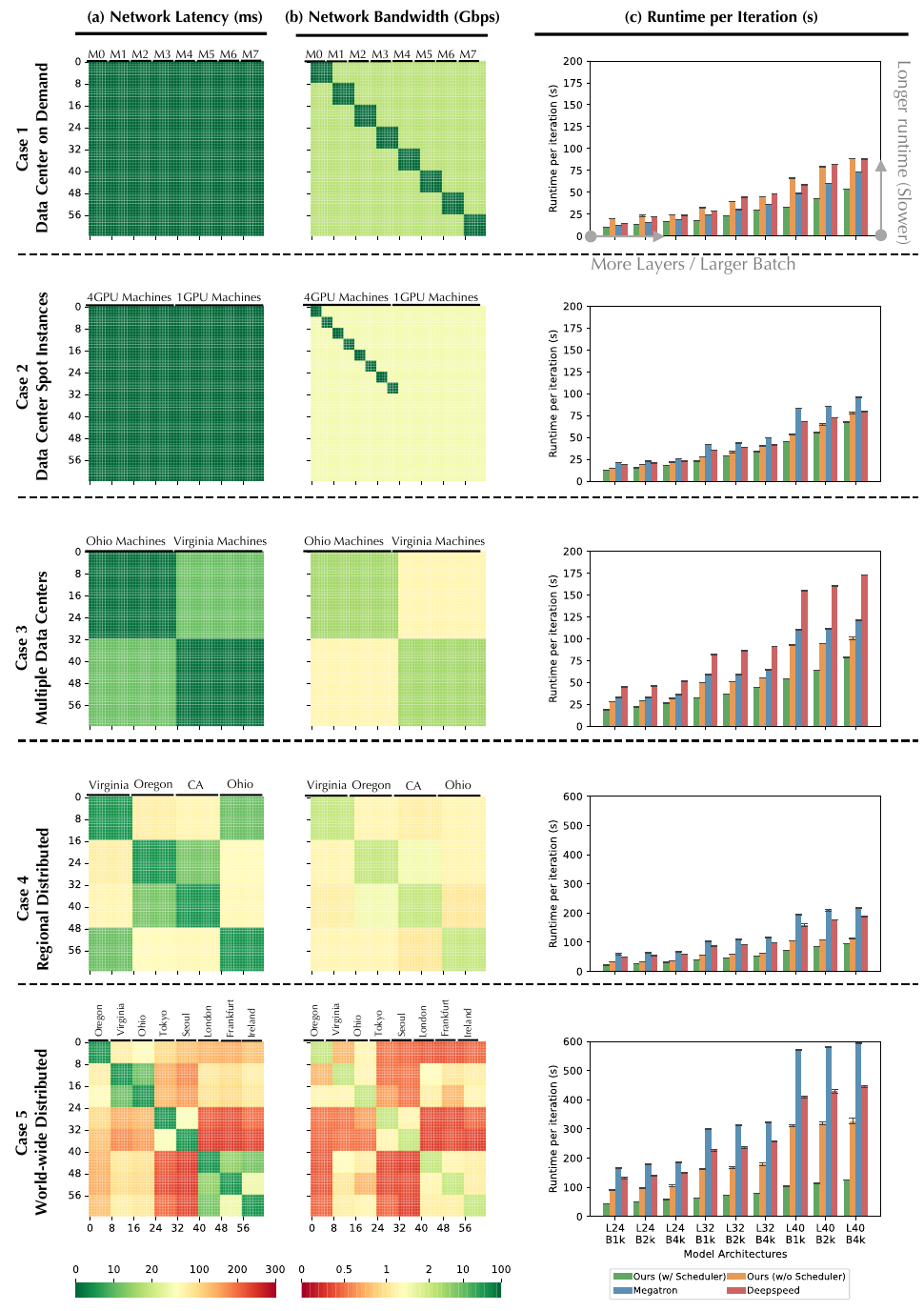}
    \caption{End to end compassion of in terms of \textit{runtime of each iteration} in 5 different scenarios. We illustrate the comparison of Megatron, Deepspeed and our system with and without scheduler.}
    \label{fig:exp_appendix}
\end{figure}

In Figure \ref{fig:exp_appendix}, we plot \textit{runtime of each iteration} for each scenario; this is a supplement to Figure \ref{fig:exp}.

\subsection{Deployment on FluidStack}

We believe that having more realistic measurements and an end-to-end run can provide more pervasive statements for decentralized training. To this end, we conducted an additional experiment.

We rent 32 A40 GPUs (each with 48GB GPU memory, and 149.7 peak FP16 TFLOPS) from FluidStack~\cite{fluidstak}, which consists of a group of geo-distributed GPU clusters, in (1) US Mid and (2) US East. We get the communication delay and bandwidths between GPUs as below:

\begin{itemize}
    \item Intra-US Mid: delay $0.5_{\pm0.1}$ ms; bandwidth $10.40_{\pm1.11}$ Gbps;
    \item Intra-US East: delay $0.5_{\pm0.1}$ ms; bandwidth $11.98_{\pm1.92}$ Gbps;
    \item US Mid to US East: delay $21.8_{\pm0.3}$ ms; bandwidth $3.87_{\pm1.07}$ Gbps;
    \item US East to US Mid: delay $21.8_{\pm0.3}$ ms; bandwidth $3.73_{\pm1.38}$ Gbps.
\end{itemize}

\begin{figure}[t!]
    \centering
    \includegraphics[width=0.7\linewidth]{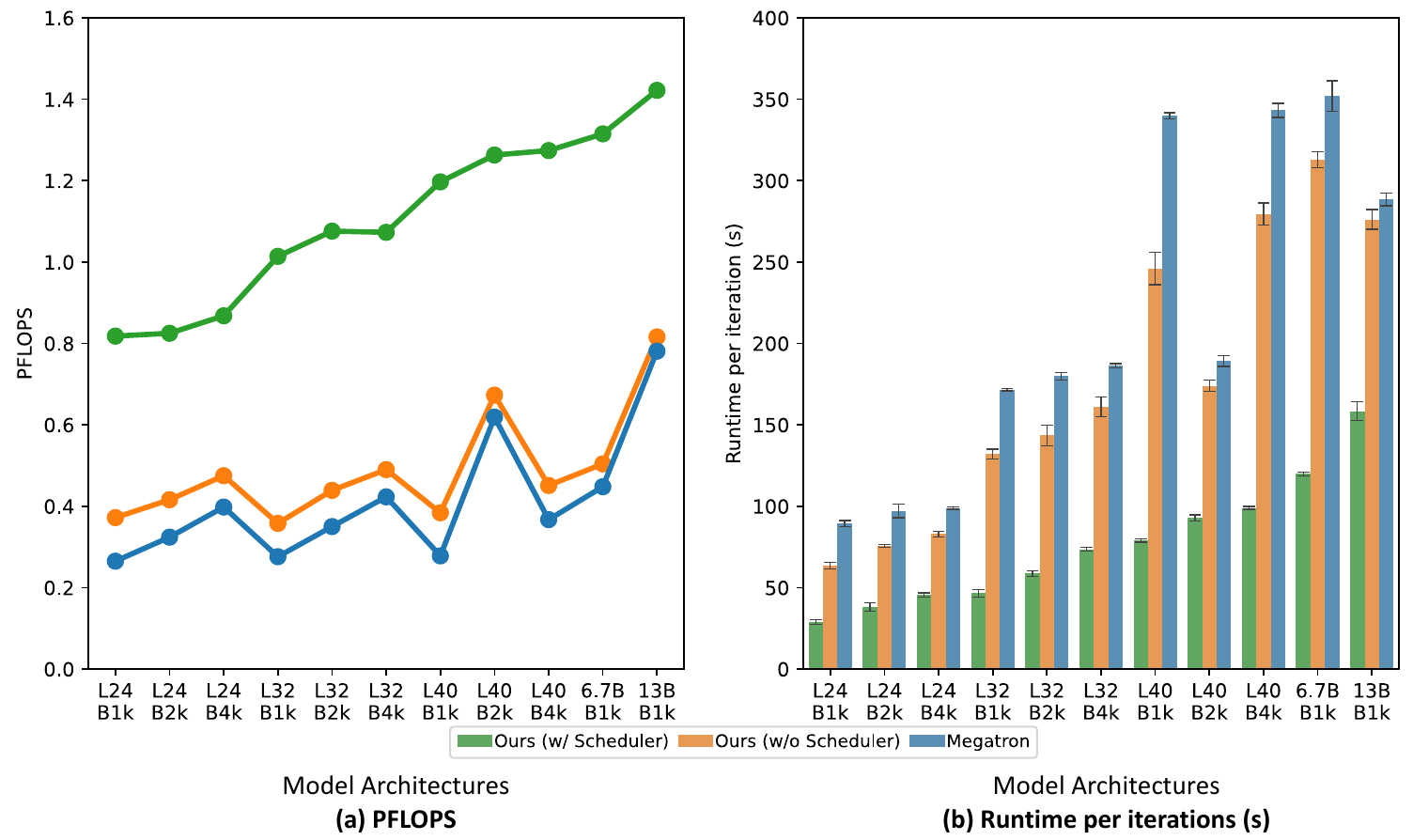}
    \caption{End to end compassion of in terms of the cluster's whole PFLOPS in (a) and runtime of each iteration in (b). We illustrate the comparison of Megatron and our system with and without scheduler.}
    \label{fig:real_run}
\end{figure}

We conducted an end-to-end run of the same training task of GPT3-1.3B without artificially controlling the bandwidth and latency. We also explore the training tasks of larger scale GPT3 models, including GPT3-6.7B, and GPT3-13B with a batch size of 1024.  The performance in terms of the total number of floating point operations per second (PFLOPS) and runtime of each iteration are illustrated in Figure \ref{fig:real_run}. This is a promising result of decentralized training --- for GPT3-1.3B model with 40 layers and 4K batch size, we archive $27.4\%$ of the peak FLOPS of the cluster, for GPT3-6.7B and GPT3-13B, we obtain $26.4\%$ and $29.7\%$ respectively.

\end{document}